\documentclass[twocolumn,superscriptaddress,nofootinbib,aps,prl,floatfix,preprintnumbers,amsmath,amssymb,groupedaddress]{revtex4}

\usepackage{dsfont}
\usepackage{epsfig}
\usepackage{slashed}
\usepackage{bbold}
\usepackage{psfrag}
\usepackage{color}
\PassOptionsToPackage{caption=false}{subfig}
\usepackage{subfig}
\usepackage{multirow}
\usepackage{booktabs}
\usepackage{hyperref}

\begin{document}
%%%%%%%%%%%%%%%%%%%%%%%%%%%%%%%%%%
%%%%%%%%%%%%%%%%%%%%%%%%%%%%%%%%%%
\title{Closing the stop gap}
%%%%%%%%%%%%%%%%%%%%%%%%%%%%%%%%%%
%%%%%%%%%%%%%%%%%%%%%%%%%%%%%%%%%%
\author{Michal Czakon}
\affiliation{Institut f\"ur Theoretische Teilchenphysik und Kosmologie,
RWTH Aachen University, D-52056 Aachen, Germany}

\author{Alexander Mitov}
\affiliation{Cavendish Laboratory, University of Cambridge, Cambridge CB3 0HE, UK}

\author{Michele Papucci}
\affiliation{Theoretical Physics Group, Lawrence Berkeley National
  Laboratory, Berkeley CA 94720}
\affiliation{Department of Physics, University of California, Berkeley
CA 94720}

\author{Joshua T. Ruderman}
\affiliation{Theoretical Physics Group, Lawrence Berkeley National
  Laboratory, Berkeley CA 94720}
\affiliation{Department of Physics, University of California, Berkeley
  CA 94720}
  \affiliation{Center for Cosmology and Particle Physics, \\Department of Physics, New York University, New York, NY 10003}
\author{Andreas Weiler}
\affiliation{DESY, Notkestrasse 85, D-22607 Hamburg, Germany}
\affiliation{Theory Division, CERN, 1211 Geneva 23, Switzerland}

%\pacs{PACS}
%\keywords{keywords}
\preprint{CERN-PH-TH-2014-114, DESY 14-107, Cavendish-HEP-14/04, TTK-14-12}
\begin{abstract}
Light stops are a hallmark of the most natural realizations of weak-scale supersymmetry. While stops have been extensively searched for, there remain open gaps around and below the top mass, due to similarities of stop and top signals with current statistics. We propose a new fast-track avenue to improve light stop searches for R-parity conserving supersymmetry, by comparing top cross section measurements to the theoretical prediction. Stop masses below $\sim 180\,{\rm GeV}$ can now be ruled out for a light neutralino. The possibility of a stop signal contaminating the top mass measurement is also briefly addressed.
\end{abstract}
\maketitle
%%%%%%%%%%%%%%%%%%%%%%%%%%%%%%%%%%%%%%%%%%%%%%

{\it\bf Introduction:} 
One of the open questions in particle physics is why the weak and gravitational forces have such different strengths. If this 
\emph{hierarchy problem} has a solution dictated by microscopic dynamics, one expects new particles not far from the weak scale, 
${\cal O}(100\,{\rm GeV})$, in the form of partners of the Standard Model (SM) particles, responsible for insulating the Higgs mass 
from large ultraviolet quantum corrections. Weak-scale supersymmetry (SUSY) is a leading candidate for such a
microscopic solution of the hierarchy problem and the mechanism is most natural if the partners of the SM particles having the 
largest coupling to the Higgs field are light~\cite{Dimopoulos:1995mi,Cohen:1996vb}, the top squark being the most prominent one. This region of the SUSY parameter space has 
been called Natural SUSY in recent years~\cite{Papucci:2011wy}.  Many theoretical studies~\cite{Brust:2011tb,Kats:2011it,Drees:2012dd,Plehn:2012pr,Alves:2012ft,Han:2012fw,Kilic:2012kw,Graesser:2012qy,Krizka:2012ah,Delgado:2012eu,Bai:2013ema,Boughezal:2013pja,Han:2013lna,Papucci:2014rja} and experimental searches~\cite{Aad:2014bva,Aad:2014mha,Aad:2014qaa,Aad:2013ija,Khachatryan:2014doa,Chatrchyan:2013mya,Chatrchyan:2013xna,Chatrchyan:2013xsw,Chatrchyan:2013lya,ATLAS-CONF-2014-014,ATLAS-CONF-2013-068,ATLAS-CONF-2013-048,ATLAS-CONF-2013-037,ATLAS-CONF-2013-061,CMS-PAS-SUS-13-018,CMS-PAS-SUS-13-009,CMS-PAS-SUS-13-004,CMS-PAS-SUS-13-015} aimed at probing Natural SUSY models have therefore focused on searches for the top (and bottom) squarks $\tilde t$~($\tilde b$). 

In R-parity conserving scenarios, current LHC limits reach up to about $700\,{\rm GeV}$, depending  on the value of the lightest SUSY 
particle (LSP) mass, usually taken to be a neutralino ($\chi^{0}_{1}$) or a gravitino ($\tilde G$). However, unconstrained regions for 
lighter values of stop masses still remain, the most important being the one where $m_{\tilde t} \sim m_{t} \gg m_{\chi^{0}_{1},\tilde G}$ and $\tilde t$ decays into (off-shell) top and the LSP, {\emph i.e.}~where $\tilde 
t$ decays are kinematically very similar to top decays. Given that the production cross section for top squarks is much smaller than 
the one for top quarks ($\sigma_{\tilde t} \sim 0.15\,\sigma_{t\bar t}$ for $m_{\tilde t}\sim m_{t}$ at the LHC), constraining these 
\emph{stealth stop} models~\cite{Fan:2012jf,Fan:2011yu,Csaki:2012fh} is particularly challenging. All of the strategies studied in the literature focused on exploiting the 
subtle kinematical differences between the top and stop production and/or decays
~\cite{Han:2012fw,Kilic:2012kw,Han:2013lna}. Furthermore, the best known discriminating 
kinematical variables, such as the lepton rapidity distribution or the dilepton angular correlations, are either plagued by large theoretical and pdf uncertainties 
or require very large statistics, 
only accessible in future LHC runs~\cite{Agashe:2013hma}. To date, the strongest constraints come from dedicated searches using multivariate analyses and provide only 
a partial exclusion of the stealth stop window~\cite{Chatrchyan:2013xna,Aad:2014qaa}. 
Open gaps remain. For instance,  for massless neutralino, $80\,{\rm GeV} \lesssim m_{\tilde t} \lesssim 100\,{\rm GeV}$ or  $m_{\tilde t}$ around $m_{t}$ are still allowed. While model-dependent limits in these gaps arise from indirect Higgs couplings constraints (see \emph{e.g.}~\cite{Espinosa:2012in,Gori:2013mia,Grojean:2013nya,Schlaffer:2014osa,Fan:2014txa}) and from $\tilde t \rightarrow c \,\chi_{0}^{1}$ searches~\cite{ATLAS-CONF-2013-068,CMS-PAS-SUS-13-009}, we stress that no robust exclusion is currently available.  

\begin{figure*}[tb]
\centering
\includegraphics[width=\linewidth]{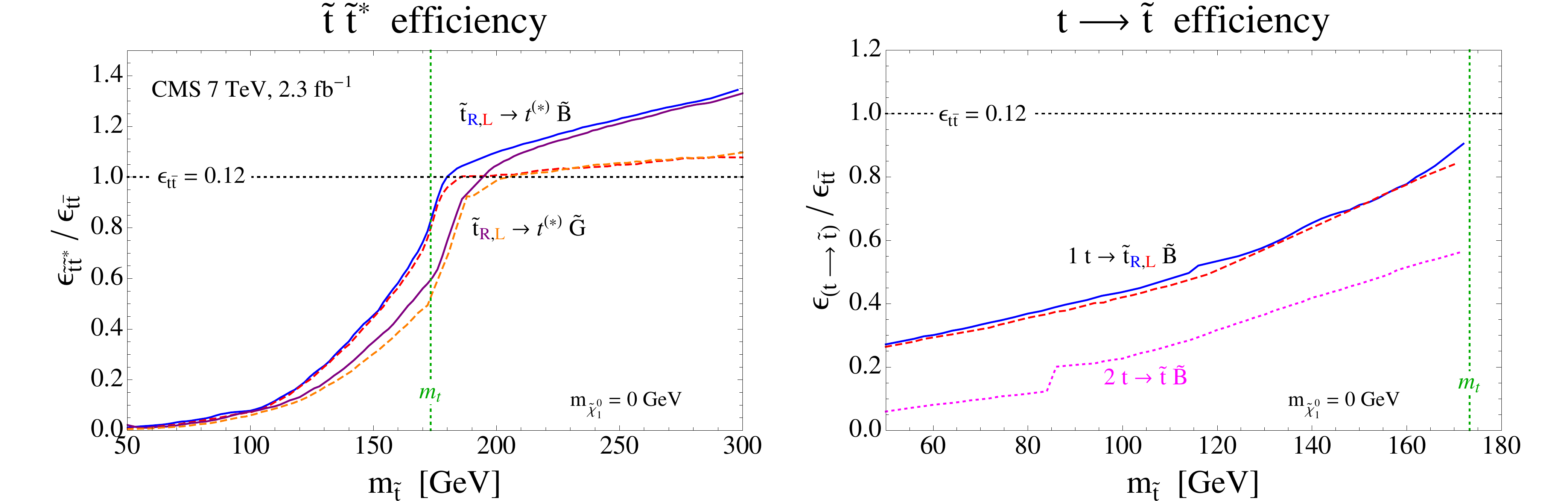}
\caption{Efficiencies and acceptances for stop pair production (left) and top pair production with one or two tops decaying to stop 
and neutralino (right) in the CMS top cross section measurement considered~\cite{Chatrchyan:2012bra}. The efficiencies are normalized to the SM top 
efficiency of $0.12$. Solid lines refer to a right-polarized stop (blue for the case of bino LSP, purple for the gravitino LSP), while 
dashed lines to a left-polarized stop (red for bino LSP and orange for gravitino LSP). We use Pythia for $2(t\rightarrow \tilde t)$ (dotted magenta), neglecting polarization and off-shell effects.\label{fig:effs}}
\end{figure*}

In this letter we propose a different, \emph{complementary} approach for constraining light top squarks. Instead of focusing on 
discriminating \emph{differences} between SUSY signal and SM background, our method is based on exploiting the kinematical 
\emph{similarities} between top and stops in this region. Namely, if stop production and decays are kinematically very similar to the 
SM top ones, then SUSY contributions may bias SM measurements. Similar methods have been proposed for constraining new physics with $W^{+}W^{-}$ measurements~\cite{Lisanti:2011cj,Feigl:2012df,Curtin:2012nn,Rolbiecki:2013fia,Curtin:2013gta,Curtin:2014zua,Kim:2014eva}. Therefore, we propose to use top SM measurements and SM 
theoretical predictions to set limits on the stop contamination in $t\bar t$ event samples. We will illustrate our method by focusing on  
one of the most inclusive top properties, the top production cross section, $\sigma_{t\bar t}$. The inclusiveness has the advantage of 
reducing theoretical uncertainties. Furthermore the theoretical prediction for $\sigma_{t\bar t}$ in the SM~\cite{Nason:1987xz,Beenakker:1988bq} has been recently improved to NNLO+NNLL by a multi-year effort of two of the authors~\cite{Baernreuther:2012ws,Czakon:2012zr,Czakon:2012pz,Czakon:2013goa,Cacciari:2011hy}, providing \cite{Czakon:2013goa,Czakon:2011xx}  
$\sigma_{t\bar t}^{LHC7} = 172^{+4.4}_{-5.8}{\rm(scale)}^{+4.7}_{-4.8}{\rm (pdf)}\,{\rm pb}$
for $m_{t}=173.3~{\rm GeV}$. Interestingly, the theoretical uncertainties are now comparable to 
the experimental ones, providing a unique opportunity for performing this analysis: further experimental improvements alone will 
only marginally change the constraining power of this method.

{\it\bf Procedure:} In practice, in the presence of a SUSY contamination, the measured cross section is 

\begin{equation}
\sigma_{t\bar t}^{exp} = \sigma_{t\bar t}(m_{t})\left(1+\frac{\epsilon_{\tilde t  \tilde t^*}(m_{t},m_{\tilde t},m_{\chi^{0}_1})}{\epsilon_{t\bar t}(m_{t})}
\frac{\sigma_{\tilde t \tilde t^*}(m_{\tilde t})}{\sigma_{t\bar t}(m_{t})}\right) \label{eq:master}
\end{equation}
where with $\epsilon$ we collectively denote the efficiency and acceptances for an event to be selected by the experimental 
analysis. We keep the explicit mass dependence of the various quantities, and for simplicity  we include only the top squark pair production contribution.  This formula gets further modified if the top is kinematically allowed to decay to a stop, as described below.  Note that throughout this paper, we assume the stop always decays to a lighter neutralino, leaving the possibility of decays to charginos for future work.

For $m_{\tilde t} \sim m_{t}$, $\sigma_{\tilde t \tilde t^*}\sim 26\,{\rm pb}$ at $\sqrt s = 7$~TeV. Taking the efficiencies $
\epsilon_{t \bar t, \tilde t \tilde t^{*}}$ to be the same, and adding the theoretical and experimental uncertainties in quadrature, one naively expects 
to set upper bounds at $95\%\,{\rm C.L.}$ on $\sigma_{\tilde t \tilde t^{*}}$ of $45\,{\rm pb}$ and $25\,{\rm pb}$ by using the SM NLO+NLL  and NNLO+NNLL predictions for $\sigma_{t \bar t}$ respectively. This clearly indicates that it was not possible~\cite{Kats:2011it} to use our proposed method before the NNLO results were available. A similar result persists in a more careful analysis~\cite{talks} as illustrated below.

\begin{figure*}[ht]
\centering
\includegraphics[width=\linewidth]{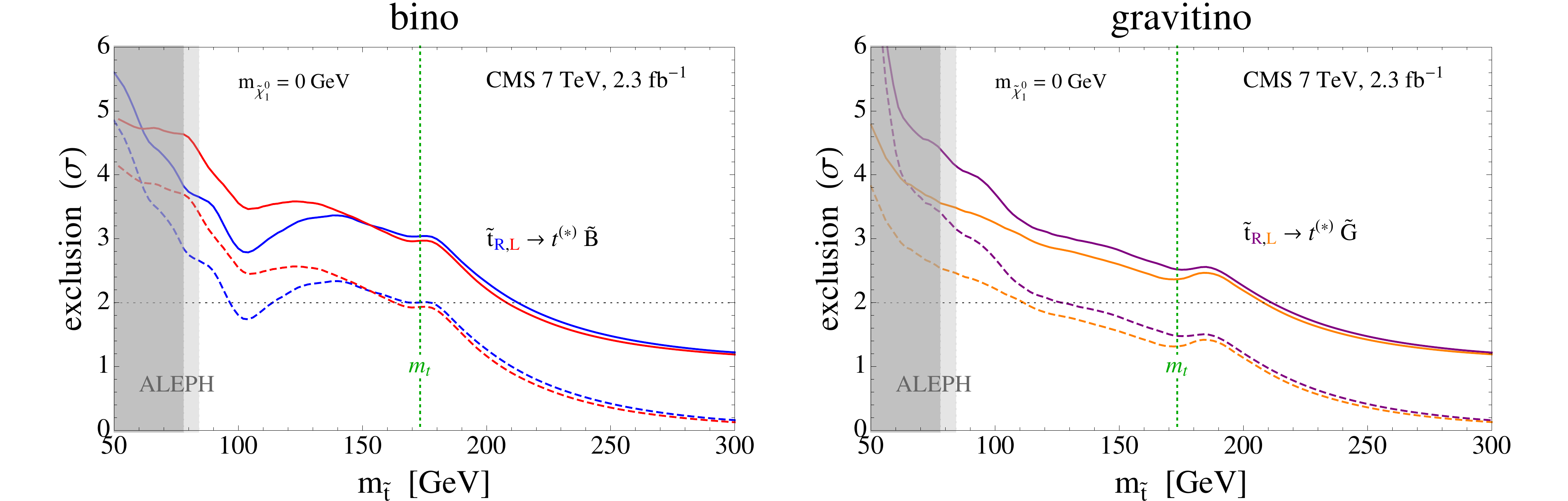}
\caption{Exclusion limits for stop decaying into a massless LSP, for bino (left) and gravitino (right). Left and right stop polarization 
are shown with (red, blue) and (orange, purple) lines respectively. Solid lines correspond to the observed limits while dashed lines 
correspond to the expected limits. LEP exclusions from ALEPH~\cite{Heister:2002hp} are shown as shaded gray (the case for minimal and maximal stop coupling to the $Z$~boson are shown).\label{fig:lim1D}}
\end{figure*}

We now describe our method in detail. For concreteness we focus on the CMS $7\,{\rm TeV}$ $2.3\,{\rm fb}^{-1}$ 
cross section measurement~\cite{Chatrchyan:2012bra}, based on dileptonic $t\bar t$ final states and using a cut and count approach, providing a measurement uncertainty $\delta\sigma_{t\bar t}/\sigma_{t\bar t}\sim 4.5\% $, comparable to the most precise LHC measurements. It is useful for illustrating our method, since, contrary to those analyses
based on multivariate (MVA) techniques, it allows us to reproduce fairly well its results without a detailed knowledge of the unpublished inner workings of the analysis. Moreover, cut and count analyses tend to be more inclusive than MVA ones and therefore they may accept a larger fraction of the contaminating SUSY signal. We stress that ultimately the study proposed here should be performed directly by the experimental collaborations. 

In the following we first discuss the case where the SM top mass is known and use $m_t = 173.3$~GeV. This assumes that a possible stop 
contamination in the $t\bar t$ sample does not bias current top mass measurements. We leave the investigation of this question to 
future work~\cite{topmass}, while we limit ourselves to showing its implications by relaxing this assumption later in this letter.

The quantity in~(\ref{eq:master}) that needs to be estimated is $\epsilon_{\tilde t  \tilde t^*}/\epsilon_{t\bar t}$. For this purpose we generated events with MadGraph~5~\cite{Alwall:2011uj}, showered and hadronized with Pythia~6.4~\cite{Sjostrand:2006za}, and performed jet clustering using FastJet~3.0~\cite{hep-ph/0512210,Cacciari:2011ma}. Both off-shell and on-shell decays of the top and stop have been properly included. In particular we find that off-shell effects are important also for the region $m_{\tilde t} > m_{t}$. We have implemented the CMS analysis in 
the ATOM package~\cite{atom} and validated it with the information provided in the experimental paper. We find very good agreement  comparing the $\bar t t$ acceptance $\times$ efficiency, see Table~\ref{tab:atom-results}.
Additional cross checks have been performed with PGS4~\cite{pgs4}.

\begin{table}[ht]
\begin{center}
\renewcommand{\arraystretch}{1.2}
\begin{tabular}{llll}
\toprule
         & $e^+e^-$ & $ \mu^+ \mu^-$ & $ e^\pm \mu^\mp$\\
\hline
{\sc atom}                               & $0.262 \pm 0.007 $         & $0.289 \pm 0.008 $              & $0.937 \pm  0.013$\\
{\sc atom} $\times$ tr. $\times$ eff.  & $0.202 \pm 0.006 $         & $0.274 \pm 0.007 $             & $0.832 \pm 0.012 $\\
{\sc cms}                                & $0.20 \pm 0.01$         & $0.27 \pm 0.01$             & $0.84 \pm 0.04$\\
\bottomrule
\end{tabular}
\caption{Comparison of the $\bar t t$ acceptance $\times$ efficiency $\times$ branching ratio (\%) between CMS and ATOM after event selection and application of one b-tag. The first line is the pure ATOM result assuming 100\% efficient electron/muon reconstruction and triggering. The second line is the ATOM result multiplied by the average of the ranges for these efficiencies as quoted in the CMS paper~\cite{Chatrchyan:2012bra}, and the third line is the CMS result from the same paper. We show the statistical MC error of the ATOM result and the error quoted by CMS, respectively.
\label{tab:atom-results}}
\end{center}
\end{table}%

To further reduce the recasting uncertainties, we will always use the ratio $\epsilon_{\tilde t  \tilde t^*}/\epsilon_{t\bar t}$ with both $\epsilon$'s estimated with the same tools. We use the NLO+NLL expression for the 
stop cross section~\cite{Beenakker:1997ut,Beenakker:2010nq,Kramer:2012bx} and neglect SUSY effects in the top production cross section~\cite{Sullivan:1996ry,Hollik:1997hm} since they are 
negligible for the spectrum considered here. Our findings are shown in Fig.~\ref{fig:effs}a for a massless lightest SUSY particle 
(LSP). The efficiency for stop pair production relative to top quickly drops for $m_{\tilde t}<m_{t}$, but it is still sizable for $m_{\tilde t}
\sim 100\,{\rm GeV}$, while it increases for $m_{\tilde t}> m_{t}$. We consider both the case of stop decaying into bino and 
gravitino LSP, and the case of different polarization in stop decays, by presenting pure $\tilde t_{L}$ and $\tilde t_{R}$ cases.
Differences between bino and gravitino LSP are most significant in the region where $m_{\tilde t}\sim m_{t}$ while stop polarization 
greatly affects the efficiency in $m_{\tilde t}>m_{t}$ region. For the bino LSP case, when $m_{\tilde t}+m_{\chi^{0}_{1}}<m_{t}$, the decays $t
\rightarrow \tilde t \chi^{0}_{1}$ are open. In this case eq.~(\ref{eq:master}) gets modified as
\begin{widetext}
\begin{equation}
\sigma_{t\bar t}^{exp} = \sigma_{t\bar t}(m_{t})\left((1-B)^{2}+2B(1-B)\frac{\epsilon_{t\bar t, \,1 (t\rightarrow \tilde t)}}{\epsilon_{t\bar t}}+B^{2} 
\frac{\epsilon_{t \bar t, \, 2 (t \rightarrow \tilde t)}}{\epsilon_{t\bar t}} + \frac{\epsilon_{\tilde t \tilde t^{*}}}{\epsilon_{t\bar t}}
\frac{\sigma_{\tilde t \tilde t^*}}{\sigma_{t\bar t}}\right) 
\end{equation}
\end{widetext}
where $B$ is the branching ratio of $t\rightarrow \tilde t \chi^{0}_{1}$ which can be as large as ${\cal O}(10\%)$~\cite{Djouadi:1996pi}. For simplicity we have not made explicit the mass dependence of the various quantities. We show the behavior of $\epsilon_{t\bar t,\,1 (t\rightarrow \tilde t)}$ and $\epsilon_{t\bar t,\,2 (t \rightarrow \tilde t)}$ in Fig.~
\ref{fig:effs}b. We find that events with a single top SUSY decay provide a sizable contribution to the SUSY signal while double top 
SUSY decays are usually negligible. The kink in the purple line on the right (the $2 t\rightarrow 2 \tilde t\, 2\chi^{0}_{1}$ efficiency) is due to Pythia being used as a generator (in this specific case, for computational limitations) which does not model the transition to off-shell decays correctly.

\begin{figure*}[thbp]
\centering
\includegraphics[width=\linewidth]{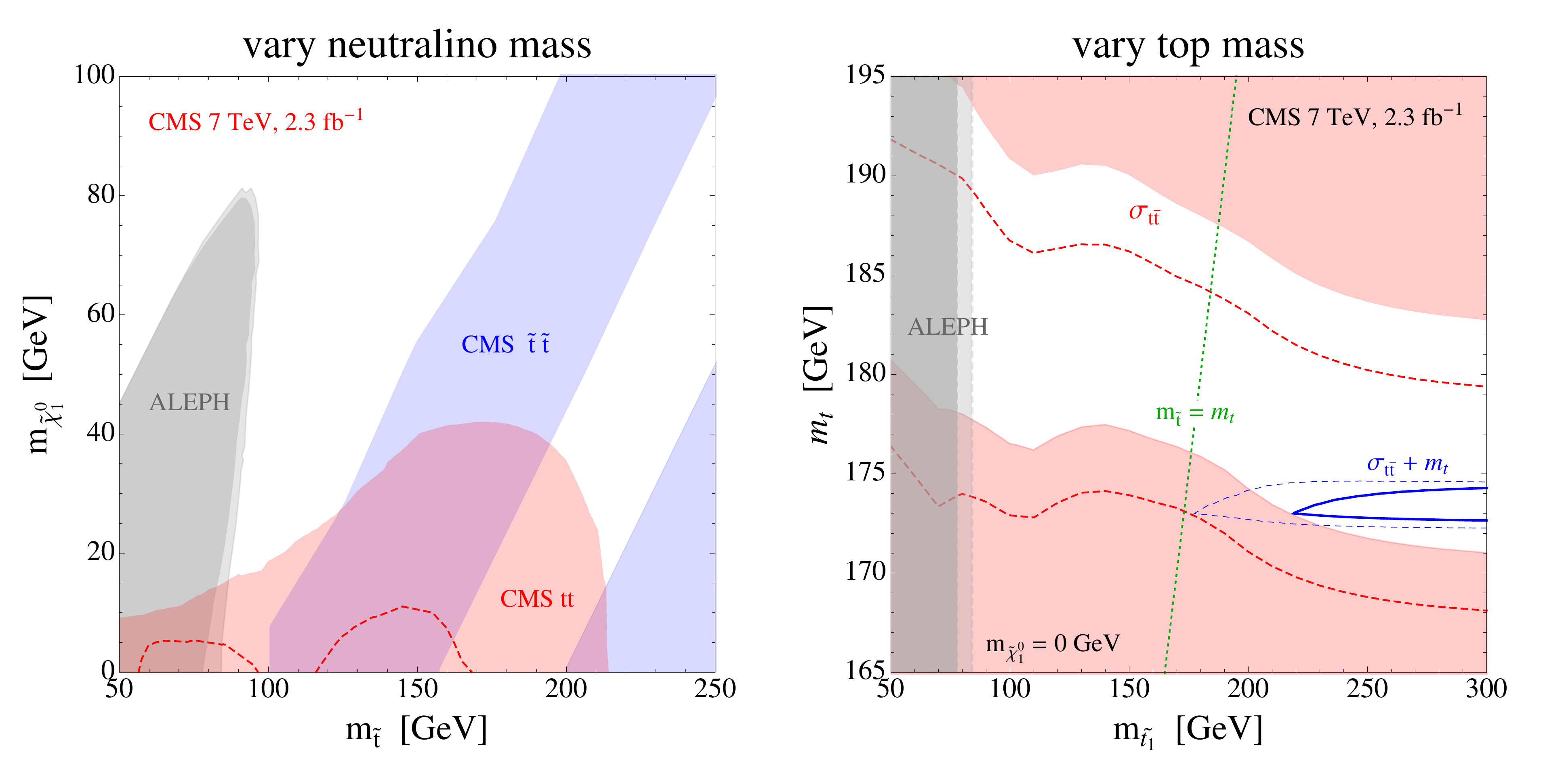}
\caption{Left: two dimensional $95\%$~C.L. exclusion limits in the neutralino-stop mass plane. Our derived limits are shown in red 
(with expected limits shown as a dashed line), LEP limits~\cite{Heister:2002hp} in gray while the CMS direct stop search in the light stop region~\cite{Chatrchyan:2013xna} is shown 
in blue. Right: excluded regions for massless neutralino in the stop-top mass plane.   Excluded region from our analysis derived using the top cross section alone (i.e. without assuming prior knowledge of the top mass) are shaded in red, while the LEP limits are shown in gray. The effect of combining the $\sigma_{t\bar t}$ measurement with current $m_{t}$ measurements (assuming no stop contamination) is shown as a blue line. Expected limits are shown as dashed lines.  For both plots we assume right-handed stop,  $\tilde t_{R}$.\label{fig:lim2D}}
\end{figure*}

{\bf Results:} We first present the limits for negligible LSP masses. Both the cases of bino and gravitino LSP are illustrated in Fig.~
\ref{fig:lim1D}. For simplicity, to set limits on the size of the SUSY signal, we have used a $\chi^{2}$ approximation, including signal and background errors, and combining errors in quadrature whenever necessary. The measured top production cross section by the analysis we considered lies below the current SM prediction, 
thus strengthening the stop limits. To provide a better sense on the power of this method with the current experimental and 
theoretical uncertainties, we also present (as dashed lines) the expected limits. 

We find that our approach is able to extend ALEPH limits~\cite{Heister:2002hp} beyond the LEP kinematical 
range into a region currently unconstrained by LHC direct searches. Stop mass limits based on the top cross section may 
reach and extend beyond the top mass, with the bino LSP case being more strongly constrained at higher stop masses and being 
less constrained, for $\tilde t_{R}$ decays around $80-100\, {\rm GeV}$, due to the less efficient $t
\rightarrow \tilde t \chi^{0}_{1}$ decays, see Fig.~\ref{fig:effs} (right).

In Fig.~\ref{fig:lim2D}a we present the case where the bino mass is allowed to move in the $(m_{\tilde t},\, m_{\chi^{0}_1})$ plane,  
comparing our limits to those obtained by other existing direct stop searches~\cite{Heister:2002hp,Chatrchyan:2013xna}. Our method is closing the stealth stop window for low neutralino 
masses, $m_{\chi^{0}_1}\lesssim 20\,{\rm GeV}$, while it is not effective for higher masses because signal rates rapidily become too low 
with increasing $m_{\chi^{0}_1}$. 

Finally, in Fig.~\ref{fig:lim2D}b we consider the case where the assumption 
of a known top mass is relaxed. We use the $m_{t}$ dependence of $\sigma_{t \bar t}$ presented in~\cite{Czakon:2013goa}. We show the limits of this scenario in the $(m_{\tilde t},m_{t})$ plane for massless bino. If $m_{t}
$ is not known, either due to stop contamination or to theoretical uncertainties~\cite{Juste:2013dsa}, an increase in $m_{t}$ can reduce $\sigma_{t\bar t}$, thus compensating the effects of the extra SUSY contributions. Therefore the top cross section is now allowing a significantly larger band in the top--stop mass plane. However a $10\,{\rm GeV}$ shift in the top mass is required to re-open the stop window all the way below $150\,{\rm GeV}$. While this shift is likely too large to be 
allowed by current top mass measurements  given the agreement across different analysis techniques and given the ${\cal O}(2\,{\rm GeV})$ uncertainty on $m_{t}$ in the endpoint analysis in~\cite{CMS-PAS-TOP-11-027}, the precise extent of the allowed regions can ultimately be constrained only by studying SUSY contamination in top mass analyses.  In Fig.~\ref{fig:lim2D}b we also show the limit that would be achieved by combining the cross section measurement with a mass measurement of $m_{t}=173.34\pm 0.76\, {\rm GeV}$~\cite{ATLAS:2014wva}, in order to illustrate the sensitivity assuming present mass measurements are not significantly impacted by the presence of stops.

{\bf Discussion:} We have introduced a novel method for constraining light stops with precision top cross section measurements at 
the LHC. The idea of using precision SM measurements to constrain BSM physics is well known for indirect observables (like electroweak precision measurements or flavor violating observables), but mostly unexplored at high energy colliders, such as the LHC, where a dichotomy between 
``measurements'' and ``searches'' is often present.
This type of studies can be very powerful in covering the shortcomings of standard searches, but clearly require high precision for both theory and experiment which, at present, makes them applicable only to a select {\it but growing} set of LHC observables.
Nevertheless, precision studies provide a new avenue towards 
light new physics exhibiting kinematics very similar to the SM backgrounds.

For the specific example discussed here, further improvements to our findings may be possible. First of all, on the 
experimental side, measurements of the top cross section with the full LHC Run I dataset will reduce the statistical uncertainties 
and may help with reducing the systematic uncertainties which dominate the errors in the analysis we have considered here. On the theoretical 
side, sizeable PDF uncertainties (and discrepancies among different PDF sets) may be reduced by taking ratios of production cross sections at different energies, found to be fairly insensitive to stop contamination in~\cite{Czakon:2013xaa}. Fully differential NNLO calculations, that are expected to appear in the very near future, will help further reduce the theoretical uncertainty on the predicted fiducial cross section. Ultimately this method can complement direct stealth stop searches which will become accessible with higher luminosities in the next LHC runs.

%\begin{acknowledgments}
{\it\bf Acknowledgments:} We thank Lance Dixon,  Beate Heinemann, Ian Low, Michelangelo Mangano, and Matt Reece for discussions. M.~P. and J.~T.~R. thank the CERN TH group and the Aspen Center for Physics for their hospitality. J.~T.~R. also thanks CFHEP at IHEP for hospitality. The work of M.~C. was supported by the German Research Foundation (DFG) via the Sonderforschungsbereich/Transregio SFB/TR-9 ``Computational Particle Physics", and the Heisenberg programme. The work of A.~M. is supported by the UK Science and Technology Facilities Council [grants ST/L002760/1 and ST/K004883/1]. The work of A.~M. was supported in part by ERC grant 291377 ``LHCtheory: Theoretical predictions and analyses of LHC physics: advancing the precision frontier".
M.~P. was supported in part by the Director, Office of Science, Office of High Energy Physics, of the US Department of Energy under Contract DE-AC02-05CH11231.
J.~T.~R. is supported by a fellowship from the Miller Institute for Basic Research in Science.  The work of A.~W. was supported 
in part by the German Science Foundation (DFG) under the Collaborative Research Center (SFB) 676.

%\end{acknowledgements}
%%%%%%%%%%%%%%%%%%%%%%%%%%%%%%%%%%%%%%%%%%%%%%

%%%%%%%%%%%%%%%%%%%%%%%%%%%%%%%%%%%%%%%%%%%%%%
%%%%%%%%%%%%%%%%%%%%%%%%%%%%%%%%%%%%%%%%%%%%%%
\end{document}